\documentclass[twocolumn]{revtex4-1}

\usepackage{amsmath, amsfonts,bbm}
\usepackage{graphicx}
\usepackage{subfigure}
\usepackage{color}
\usepackage{epstopdf}
\DeclareGraphicsExtensions{.eps}
\usepackage{graphicx}
\usepackage[nice, tight]{units}

\newcommand{\bra}[1]{\ensuremath{\left\langle{#1}\right\vert}}
\newcommand{\ket}[1]{\ensuremath{\left\vert{#1}\right\rangle}}
\newcommand{\hcz}{\textit{h}\textsc{cz}}

\def\>{\rangle}
\def\<{\langle}

\begin{document}

\title{Engineering integrated photonics for heralded quantum gates}

\author{T.~Meany~$^{*, \dagger}$}

\affiliation{Centre for Ultrahigh bandwidth Devices for Optical Systems (CUDOS), MQ Photonics Research Centre, Department of Physics and Astronomy, Macquarie University, NSW 2109, Australia}
\email{thomasmeany1@gmail.com\\$^{\dagger}$\emph{Authors contributed equally to this work.}}

\author{D.~N.~Biggerstaff~$^{\dagger}$}
\affiliation{Centre for Engineered Quantum Systems, \\
 Centre for Quantum Computer and Communication Technology, School of Mathematics and Physics, University of Queensland, Brisbane QLD 4072, Australia}
 
\author{M.~A.~Broome}
\affiliation{Centre for Engineered Quantum Systems, \\
 Centre for Quantum Computer and Communication Technology, School of Mathematics and Physics, University of Queensland, Brisbane QLD 4072, Australia}

\author{A.~Fedrizzi}
\affiliation{Centre for Engineered Quantum Systems, \\
 Centre for Quantum Computer and Communication Technology, School of Mathematics and Physics, University of Queensland, Brisbane QLD 4072, Australia}
 
\author{M.~Delanty}
\affiliation{Centre for Ultrahigh bandwidth Devices for Optical Systems (CUDOS), MQ Photonics Research Centre, Department of Physics and Astronomy, Macquarie University, NSW 2109, Australia}

\author{M. J. Steel}
\affiliation{Centre for Ultrahigh bandwidth Devices for Optical Systems (CUDOS), MQ Photonics Research Centre, Department of Physics and Astronomy, Macquarie University, NSW 2109, Australia}

\author{A.~Gilchrist}
\affiliation{Centre for Engineered Quantum Systems, Department of Physics and Astronomy, Macquarie University, NSW 2109, Australia}

\author{G.~D.~Marshall}
\affiliation{Centre for Quantum Photonics, HH Wills Physics Laboratory, Tyndall Avenue, Bristol BS8 1TL, UK}

\author{A.~G.~White}
\affiliation{Centre for Engineered Quantum Systems, \\
 Centre for Quantum Computer and Communication Technology, School of Mathematics and Physics, University of Queensland, Brisbane QLD 4072, Australia}

\author{M.~J.~Withford}
\affiliation{Centre for Ultrahigh bandwidth Devices for Optical Systems (CUDOS), MQ Photonics Research Centre, Department of Physics and Astronomy, Macquarie University, NSW 2109, Australia}

\begin{abstract}
Scaling up linear-optics quantum computing will require multi-photon gates which are compact, phase-stable, exhibit excellent quantum interference, and have success heralded by the detection of ancillary photons. We investigate implementation of the optimal known gate design which meets these requirements: the Knill controlled-$Z$ gate, implemented in integrated laser-written waveguide arrays. We show that device performance is more sensitive to the small deviations in the coupler reflectivity, arising due to the tolerance values of the fabrication method, than phase variations in the circuit. The mode fidelity was also shown to be less sensitive to reflectivity and phase errors than process fidelity. Our best device achieves a fidelity of $0.931{\pm} 0.001$ with the ideal $4 {{\times}} 4$ unitary circuit and a process fidelity of $0.680{\pm}0.005$ with the ideal computational-basis process.
\end{abstract}
\maketitle

\section{Introduction} 
Recent advances in on-chip integration of efficient photon sources~\cite{Schell2013,Silverstone2013a,Meany20014,gazzano_qdcnot} and detectors~\cite{Calkins2013a,Reithmaier2013,Sahin2013} hold promise for significant progress in the linear optics architecture for quantum information processing and simulation. This architecture relies on the insight that entangling quantum gates can be realised probabilistically by interacting photonic qubits using effective optical nonlinearities induced by measurement~\cite{Knill2001a}.
However, efficient scaling in this architecture necessitates entangling gates which are both logically and physically scalable. Logical scalability requires successful operation to be heralded nondestructively, typically by the detection of additional `ancilla' photons. 
Physical scalability, meanwhile, requires a compact and phase-stable architecture.
Integrated, non-heralded entangling gates have been demonstrated~\cite{Politi2008, Crespi_CNOT_2011} as have heralded gates in bulk optics~\cite{FransonGate,okamoto2011}. 
However, gates which are scalable in both the logical and physical senses have not been implemented to date due to their geometric complexity as well as the need for low circuit losses.

The simplest heralded entangling two-qubit photonic gate design, with the highest known success probability, was found by Knill~\cite{E_Knill_PRA_C_Phase} and implements a controlled-$Z$ operation with probability $\nicefrac[]{2}{27}$. This heralded \textsc{cz} design, henceforth called the \hcz{}, relies on pairwise non-classical interference of four indistinguishable photons in a circuit with four particular beamsplitters (BSs), as shown in Fig.~
1(a), as well as a stable phase shift of precisely $\pi$ between the first and second BS pairs. 

Integrated arrays of coupled waveguides could enable compact, phase-stable circuits with the requisite splittings and phase for Knill's design. However, integrated \hcz{} circuits require a physical swapping of neighboring modes; such crossovers are difficult to achieve lithographically. Laser-written waveguides have recently been used to demonstrate a wide variety of quantum photonics circuitry~\cite{Grahamoptex,Smith:09}. The femtosecond-laser direct-write technique in particular allows 3D waveguide arrays, simplifying waveguide crossovers, along with demonstrated high mode indistinguishability~\cite{Grahamoptex}, 
 and has recently been employed for multiport and arbitrary-phase directional couplers~\cite{Meany_multiport,spagnolo2013three,heilmann_phase}, all-optical routers~\cite{keil2011all}, circuits for small-scale quantum simulations~\cite{di2013einstein,Tillmann2013,crespi2013}, quantum walks~\cite{OwensQW}, and non-heralded quantum gates~\cite{Crespi_CNOT_2011}.

Here we investigate implementation of the \hcz{} gate both theoretically and experimentally using the direct-write technique, with a particular focus on its action in the presence of deviations from optimal phase and reflectivity parameters. We derive the variation in gate performance as quantified by two metrics---the optical-circuit mode-fidelity and the computational-basis process-fidelity---with respect to such deviations. We further detail the fabrication of 12 prototype circuits including a novel and simple method for achieving the requisite internal phase, and their full characterisation using coherent techniques as well as quantum interference which confirms their excellent mode indistinguishability and suitability for the single-photon regime.

\section{Materials and Methods}

The optical circuit for the \hcz{} gate is shown in Fig.
1(a). The \emph{control} and \emph{target} qubits are each encoded as single photons across a pair of modes, labelled $C$ and $C_0$ for the control, and $T$ and $T_0$ for the target. $C_0$ and $T_0$, which encode $\ket{0}$ for their respective qubits, interact with neither the $\ket{1}$-modes nor the ancillas; our fabricated circuits contain only the four interacting modes as shown in Fig. 1(b).
Modulo local phases on the input and output modes, this circuit implements a heralded \textsc{cz} operation: conditioned on the detection of one photon in each ancilla mode it flips the sign of the $\ket{11}$-term of an arbitrary two-qubit input state $\alpha_{00}\ket{00}{+}\alpha_{10}\ket{10}{+}\alpha_{01}\ket{01}{+}\alpha_{11}\ket{11}$.
\begin{figure*}
\label{fig:Knill gate image}
\centering
\includegraphics*[width=0.98\textwidth]{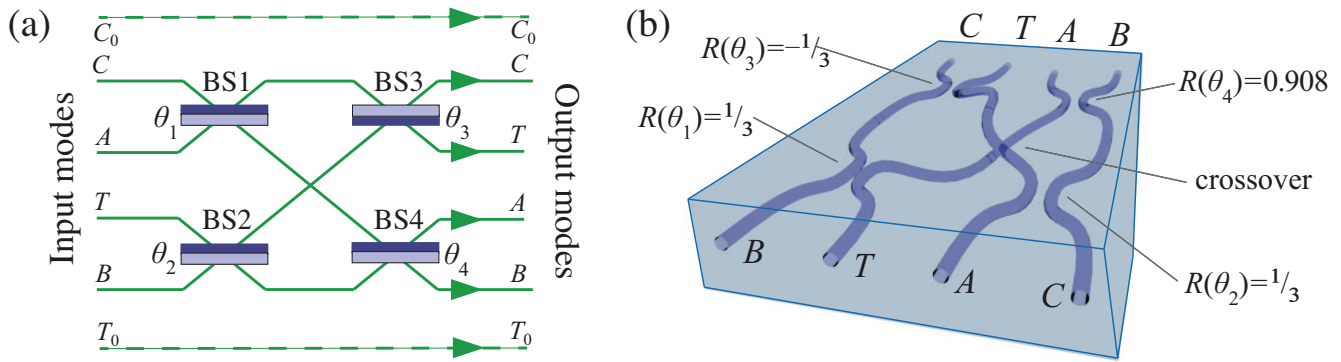}
\caption{ (a) The circuit for a \hcz{} gate showing paths for ancillary photons $A$ and $B$ as well as the computational qubits; the control (target) photon is encoded across spatial paths $C_0$ ($T_0$) representing $\ket{0}$ and $C$ ($T$) representing $\ket{1}$. The $\ket{0}$-modes do not interact in the gate; the four remaining modes undergo four beamsplitting operations with reflectivities $R(\theta_n){=}\cos^2(\theta_n)$ as described in Eq.~\eqref{eqn:beamsplitter}. The light-coloured side indicates the surface yielding a relative $\pi$ phase change upon reflection.  (b) The four interacting circuit modes modelled as a waveguide array, showing the crossover and optimal reflectivities for the BSs  implemented using evanescent coupling. 
The waveguides are separated by 127~$\mu$m at the device end facets; fan-in and fan-out regions are not shown.
The input mode labeling is reversed compared to (a) due to the reflectance of the couplers being defined as the proportion of input light which couples from one waveguide to the other.}
\end{figure*}

\subsection{Circuit modelling and design.} \label{sec:design}
 Demonstrations of entangling linear optics quantum gates relying on \emph{post-selection} of the computational photons go back more than a decade and include both free-space~\cite{obrienCNOT, Langford2005} and integrated implementations~\cite{Politi2008, Crespi_CNOT_2011}. In such gates, success relies on detecting the photonic qubits in particular output modes, precluding their use in subsequent multi-qubit operations~\cite{Knill2001a}, whereas heralded gates---where success is signalled by detection of \emph{ancillary} photons without measuring the output qubits---can be incorporated as modules in complex quantum computations. The effects of fabrication imperfections on post-selected gates have been modelled~\cite{Ralph2002,Mower14}. However such gates can effectively be simplified to include only a single instance of two-photon interference at one beamsplitter~\cite{Langford2005}; by contrast the the \hcz{}, despite its simplicity relative to other heralded gates, requires four distinct two-photon interference events at four beamsplitters. Here we quantify the effects of device imperfections on such a complex heralded gate.

We first model imperfect \hcz{} device operation as follows. We assume four single-mode waveguides coupled by BSs as in Fig.~
1(b). The quantum state of the light is described by four  bosonic creation operators $a_C^\dagger$, $a_T^\dagger$, $a_A^\dagger$, and $a_B^\dagger$ which create a  photon in the control, target, and two ancillary modes respectively. We employ the symmetric BS convention so that two modes $a_1$ and $a_2$ transform as 
\begin{widetext}
\begin{equation}
\text{BS}(\theta)\text{:} \quad a_1^\dagger \rightarrow a_1^\dagger \cos\theta {+}i a_2^\dagger \sin\theta,\;\;\; a_2^\dagger \rightarrow a_2^\dagger \cos\theta {+}i a_1^\dagger \sin\theta.
\label{eqn:beamsplitter}
\end{equation}
\end{widetext}
We first consider a photon or coherent state in a superposition of the input modes into a circuit of the form in Fig.~
1(b). Our model neglects intrinsic loss, as in our fabricated circuits such loss is nearly constant across the waveguides and therefore results only in a reduction in the output state amplitude and, thus, the gate success probability. The circuit then maps the input creation operators $a^\dagger$ to the outputs $b^\dagger$ via the transformation $b^\dagger_k{=}\sum_j U^{\text{circ}}_{jk}a^\dagger_j$, where $U^\text{circ}$ is a unitary matrix, and the mode indices $k$ and $j$ are ordered $\{C,T,A,B\}$. We allow for arbitrary splitting parameter angles $\theta_n$, $n{=}\{1,...,4\}$. The requisite internal phase shift is implemented by an additional phase of $\pi$ on BS3, equivalent to $\theta_3 \rightarrow {-}\theta_3$. Using Eq.\eqref{eqn:beamsplitter} and allowing for additional  undesired phase shifts ($a^\dagger_n\rightarrow e^{i\phi_n}a^\dagger_n$) between the BS pairs, we find that all these unwanted internal phases can be collected into a single net phase shift $\phi_N{=}\phi_c{+}\phi_a{-}\phi_b{-}\phi_t$.  For the total circuit action, modulo external local phases, we thus find:
\begin{widetext}
\normalsize
\begin{eqnarray}
U^\text{circ}{=}
	\begin{bmatrix}
	\cos{\theta_1} \cos{\theta_3} & \cos{\theta_2} \sin{\theta_3}  & \cos{\theta_1} \sin{\theta_3} & \sin{\theta_2} \sin{\theta_3} \\
	\cos{\theta_1} \sin{\theta_3} & -\cos{\theta_2} \cos{\theta_3} & \sin{\theta_1} \sin{\theta_3} & -\cos{\theta_3} \sin{\theta_2}  \\
	\cos{\theta_4} \sin{\theta_1}  & e^{i\phi_N} \sin{\theta_2} \sin{\theta_4} &-\cos{\theta_1} \cos{\theta_4}  & -e^{i\phi_N} \cos{\theta_2} \sin{\theta_4} \\
	\sin{\theta_1} \sin{\theta_4} & -e^{i\phi_N} \cos{\theta_4} \sin{\theta_2}  & -\cos{\theta_1} \sin{\theta_4}  & e^{i\phi_N} \cos{\theta_2} \cos{\theta_4} \\
	\end{bmatrix}.
	\label{eqn:Ucirc}
\end{eqnarray}
\end{widetext}
Up to external phases and in the absence of net phase $\phi_N$, the ideal matrix $U^{\text{\hcz{}}}$ given by Knill~\cite{E_Knill_PRA_C_Phase} is achieved by the target angles of $\theta_1{=}\theta_2{=}\theta_3{=}\arccos \sqrt{1/3}$, and $\theta_4{=}\arccos{\sqrt{\frac{1}{2}{+}\frac{1}{\sqrt{6}}}}$.

We employ two metrics to assess the design accuracy of a physical circuit for a \hcz{} gate. The \emph {mode fidelity} $F_m$ directly compares the $4{\times} 4$ circuit mapping matrix $U^\text{circ}$ to the ideal unitary matrix $U^\text{\hcz{}}$, and is given by the normalized Hilbert-Schmidt inner product~\cite{Nielsen}: $F_m{=}\lvert \mathrm{Tr} \{ U^{\text{\hcz{}} \dagger} U^\text{circ} \} \rvert^2{/}N^2$, where $N=4$ is the number of modes. 
This metric most closely captures the differences between the manufactured integrated device and the ideal target device, but only partially captures how the device would function with qubits since it ignores the effect of measurement heralding. For instance if the ancilla modes were swapped prior to detection, $F_m$ would decrease without any operational effect on the function of the heralded gate on the qubits, as it is irrelevant which detector detects which ancilla. The second metric directly assesses the effect of the measurement-induced optical nonlinearity in the space of the qubits, and thus will be independent of such irrelevant changes. A quantum process $\mathcal{E}$ can be represented abstractly as a quantum state $\rho_\mathcal{E}$ via the \emph{Jamiolkowski isomorphism}~\cite{jamiolkowskiiso}, and the natural figure of merit for gate quality is then the \emph{process fidelity}: $F_p {=} \mathrm{Tr}\{\rho_\text{\textsc{cz}}^\dagger \rho_\mathcal{E}\}$ which simply compares the state representing the implemented process and the state $\rho_\text{\textsc{cz}}$ representing the process for an ideal \textsc{cz} gate~\cite{Gilchrist2005,White2007}. While we would expect the two measures to be roughly similar, particularly for small imperfections in \emph{e.g.} splitting ratios, there is in general no simple relationship between them, as they are differently sensitive to imperfections.

The most direct way to calculate $\rho_\mathcal{E}$ is to consider a maximally-entangled state $\ket{\phi_\text{max}}$ between the Hilbert space on which the process acts, and another fictitious space of the same dimension. The process acts on one half of the entangled state, and the resulting total state is exactly $\rho_\mathcal{E}$. As our computational input is two qubits, the entangled state is $\ket{\phi_\text{max}}{=}(\ket{0000}{+}\ket{0101}{+}\ket{1010} {+}\ket{1111})/2$. After a \textsc{cz} operation on the first two qubits the result is $\ket{\phi_\text{\textsc{cz}}}{=}(\ket{0000}{+}\ket{0101}{+}\ket{1010}{-}\ket{1111}){/}2$, and the corresponding final state representing the process is $\rho_\text{\textsc{cz}}{=}\ket{\phi_\text{\textsc{cz}}}\!\bra{\phi_\text{\textsc{cz}}}$.

Given that each qubit comprises a photon in two modes, $\ket{\phi_\text{max}}$ involves four photons encoded across eight modes, where the fictitious additional control (target) mode has creation operator $a_{C2}^\dagger$ ($a_{T2}^\dagger$). With the addition of the two ancillary modes, the 
entangled input state is thus represented using boson creation operators as $(1{+}a_T^\dagger a_{T2}^\dagger{+}a_C^\dagger a_{C2}^\dagger{+}a_C^\dagger a_{C2}^\dagger a_T^\dagger a_{T2}^\dagger)a_A^\dagger a_B^\dagger \ket{\mathbf{0}}$ where $\ket{\mathbf{0}}$ is a multimode bosonic vacuum, and creation operators for the non-interacting logical $\ket{0}$ modes are again omitted. The circuit transforms $a_C^\dagger$, $a_T^\dagger$, $a_A^\dagger$, and $a_B^\dagger$ according to $U^\text{circ}$, and gate success is heralded by measuring exactly one photon in each ancillary mode. This measurement removes these modes and induces a \textsc{cz} on the remaining modes.

A subtle problem arises when the photonic gate is not perfectly balanced. There is then a non-zero amplitude for the states proportional to $(a_C^{\dagger})^2$ and $(a_T^{\dagger})^2$, which lie outside the qubit space and represent errors. In characterising circuit performance, we account for these errors by calculating the process fidelity against a version of $\rho_\text{\textsc{cz}}$ which is extended to include these two states but with zero support, and thus any weight on these terms will always reduce $F_p$.

Figure~
2 shows the variation of the process and mode fidelity due to a deviation in a single BS reflectivity or the net internal phase.  Both fidelity metrics are much less sensitive to small deviations in phase than in splitting ratios.    Fig.~2 also shows two fidelity distributions each resulting from 2000 randomly-chosen gate simulations with simultaneous deviations in all five reflectivity and phase parameters. Perhaps unsurprisingly the mode fidelity is far less sensitive to errors overall.
\begin{figure*}
  \begin{center}
    \includegraphics*[width=0.92\textwidth]{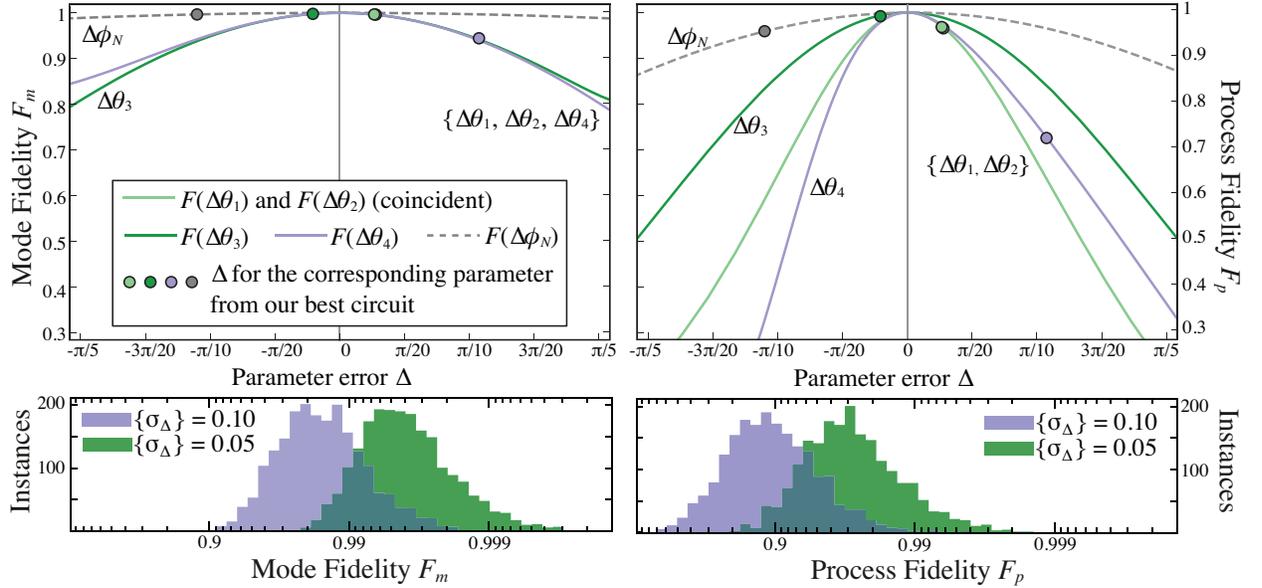}
  \end{center}
  \caption{Variation of the model mode fidelity $F_m$ and process fidelity $F_p$ with deviations $\Delta$ from the ideal BS angles and internal phase shift. The ideal phase is zero and $\Delta\Phi_N$ represents any net extra phase introduced between beamsplitters. For the BSs $\Delta$ is the variation from the ideal angle; the total reflectivity will be $\cos^2(\theta_\mathrm{ideal}+\Delta\theta)$. In both cases $\Delta$ is a length variation in the physical device. The top graphs show the fidelity when one $\Delta$ parameter is varied and the rest are held at zero. The points shown represent the deviations found in our best experimentally-characterised circuit; see Fig.~
3 for further details. Note that on this scale both the curves and points for BS1 and BS2 are indistinguishable. The bottom graphs show the fidelities for 2000 simulated instances of \hcz{} gates with all $\Delta$ parameters drawn randomly from Gaussian distributions with mean 0 and standard deviation $\sigma_\Delta$. The purple distribution has $\sigma_\Delta=0.1$, which is similar in magnitude to most of the $\Delta$ parameters from our best measured circuit; the green distribution has $\sigma_\Delta=0.05$ in order to show the fidelities achievable with a modest improvement in fabrication accuracy. The resulting green (purple) distributions have means of 0.994 (0.980) for $F_m$ and 0.962 (0.863) for $F_p$. Note the logarithmic scale on the horizontal axes.
\label{fig:modelprocessfidelities}}   
\end{figure*}

\subsection{Device fabrication}
\label{sec:devfab}

The circuits were fabricated using the FLDW technique wherein a tightly focused femtosecond laser generates a localised refractive index contrast in a glass substrate. By translating the glass in (\emph{x, y, z}) with respect to the incident laser, arbitrary 3D regions of net-positive refractive index change can be produced. Our fabrication employed a titanium sapphire oscillator (800~nm centre wavelength, $<50$~fs pulse duration) with 5.1~MHz repetition rate~\cite{Jovanovic:12,Gross:12}. A telescope was used to overfill the input pupil of a $100{\times}$ oil immersion objective which focused the laser into the boro-aluminosilicate sample (Corning Eagle 2000) for writing with 66~nJ pulses at a translation speed of 1200~mm/min. The sample was subsequently annealed to obtain a more symmetric and Gaussian refractive index profile~\cite{Arriola_annealing}. This significantly improves waveguide throughput efficiency, as shown in (ref.~\onlinecite{Meany2013}). This process yields a mode field diameter of 5~$\mu$m which has excellent overlap with an 800~nm single-mode optical fibre, and gives rise to fibre-to-fibre coupling loss of only 1.8~dB for straight waveguides of length 40~mm. Input- and output-coupling accounts for the majority of this loss, while intrinsic propagation loss is below 0.2 dB/cm~\cite{Meany2013}, and importantly intrinsic losses were found to be constant across the four waveguides to within measurement error.

 The splitting ratios of the waveguide BSs---or directional couplers---can be adjusted by changing their coupling lengths. Using a symmetric phase convention, the amplitudes in an ideal directional coupler of total length $L$ with uniform coupling constant $C$ vary sinusoidally with propagation length $z$ as $a_j^\dagger \rightarrow a_j^\dagger \cos(Cz) {+}i a_k^\dagger \sin(Cz)$, where $j,k{=}\{1,2\}$ and $0\leq z\leq L$. Since the waveguides have nominally identical profiles, the reflectivity takes the simple form $R=\cos^2\gamma$, where $\gamma=\int_0^L C(z) dz$ and we allow for variation in the coupling strength $C(z)$ along the waveguide~\cite{okamoto2006}.

While specific, arbitrary phase shifts are difficult to realise precisely without active elements using FLDW~\cite{crespi2013, heilmann_phase}, adjustments in coupler length also allowed us to achieve the requisite internal phase shift of~$\pi$. Extending $L$ such that $\gamma$ goes from $\theta$ to $(2 \pi {-} \theta)$ changes the action of the splitter to $a_j^\dagger \rightarrow a_j^\dagger \cos(-\theta) {+}i a_k^\dagger \sin(-\theta)=a_j^\dagger \cos(\theta) -i a_k^\dagger \sin(\theta)$. Exploiting this identity, we implemented the requisite phase shift by lengthening BS3 from $\gamma {=} \arccos{\sqrt{1/3}}$ to $\gamma {=} ({2\pi{-}\arccos{\sqrt{1/3}}})$. For ideal couplers the relative phase is limited to $\pm \pi/2$ and the application of this technique on BS3 yields no undesired internal phase $\phi_N$, even for slight errors in~$L$. In practice $\phi_N\neq 0$ can occur due to slight variations in local waveguide profile resulting from laser power fluctuations in fabrication, as well as from small internal path length variations.

An extensive parameter study of directional couplers was completed in order to determine the optimal laser characteristics, writing algorithm, and coupling lengths for achieving the desired reflectivities and internal phase. However, the performance of couplers written according to a particular algorithm will nevertheless vary from sample to sample, depending on the precise substrate and laser characteristics at the time of fabrication. In particular, slight refractive index differences between the two waveguides constitute a significant source of deviations from intended reflectivities. Such differences yield phase mismatch which prevents full power transfer between the waveguides, an effect which becomes more pronounced as coupling length increases. Twelve separate candidate circuits were thus fabricated, both to increase the likelihood of achieving near-optimal phase and reflectivity parameters in one or more circuits, and in order to experimentally investigate the sensitivity of device operation to parameter variations. The origin of the parameter variations can be traced to the fact that the writing laser is passively mode locked and the cavity is very long in order to achieve a balance of sufficiently high pulse energy ($>$100 nJ) combined with high repetition rates (5 MHz). This cavity is 30 m long and any perturbation can have a substantial knock on effect on the resultant pulse energy at short (sub second) time scales. A change in the pulse energy by as little as 5$\%$ can result in a change in refractive index variation and directional couplers are extremely sensitive to these variations. In principle these laser fluctuations could be improved by temperature stabilisation of the cavity area or potentially operation of the cavity in-vacuuo.

\section{Results and Discussion}
\subsection{Coherent device characterisation.}
We characterised the fabricated candidate circuits using a recently-demonstrated technique~\cite{Unitary_characterisation, BroomeAA} which yields $U^\text{meas}=r^\text{meas}\mathrm{exp}(i \phi^\text{meas})$ using only single- and two-mode bright coherent states and output intensity measurements. The moduli $r^\text{meas}_{jk}$ result from intensity measurement at each output $k$ for an input at mode $j$. The phases $\phi^\text{meas}_{jk}$ are obtained as follows: a two-mode coherent state is injected into two inputs, and a relative phase between the modes is induced via continuous path-length variation in one mode using motorised translation. The output interference fringes are recorded with fast photodiodes and an oscilloscope, and the phases $\phi^\text{meas}_{jk}$ are simply the phase differences between the pairs of resulting periodic output intensity signals $\{I_j(t)\}$. However, due to experimental noise and slight variations in the phase-setting translation velocity, it was more accurate in practice to determine the unknown phases $\phi^\text{meas}_{jk}$ by subtracting the discrete Fourier transforms of the output signals. 
\begin{figure*}
\centering  
\includegraphics*[width=0.8\textwidth]{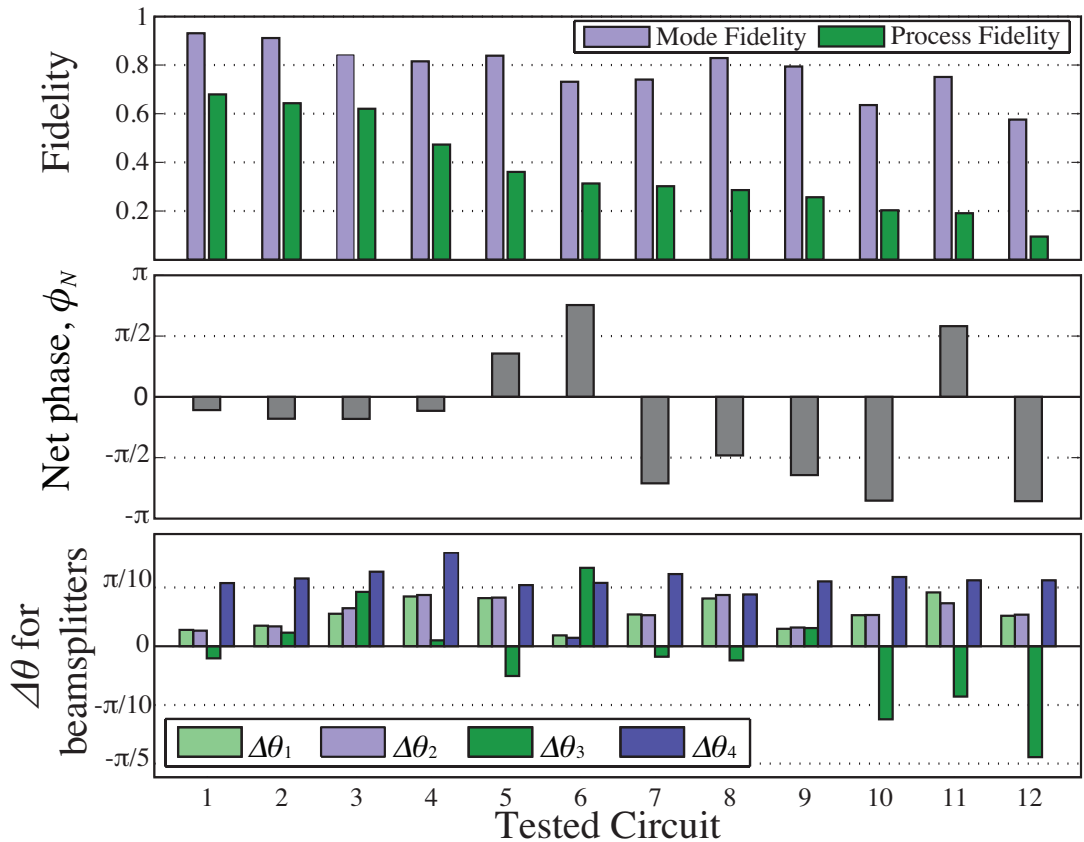}
 \caption{Results from coherent circuit characterisation. Error bars are too small to see and are thus absent. Top: Mode and process fidelity of the measured circuit mappings with the ideal \hcz{} circuit unitary, optimised over local external phases. The mean uncertainty in $F_m$ and $F_p$ are 0.002 and 0.006 respectively; these uncertainties were determined via Monte Carlo methods using the measured uncertainties in phases and moduli. Centre: Net undesired internal phase $\phi_N$. The displayed value is the mean of the four values determined from the four occurrences of $\phi_N$ in comparing $U^\text{meas}$ for each circuit to Eq.~\eqref{eqn:Ucirc}. For all 12 candidate circuits these differ by a maximum of 0.07. The variation between these four values dominated that between our many phase measurement trials for each circuit, and their standard deviation is thus taken to be our uncertainty; the mean resulting value over all 12 circuits is 0.0015. Bottom: Deviations $\Delta\theta$ from the ideal reflectivity parameters for the four BSs. For all values of $\Delta\theta$ the mean uncertainty---determined from repeated measurement trials---is 0.0026.
 In all 3 panels, the measured circuits are ordered by decreasing process fidelity.
  \label{fig:fidelity_table}
  }
\end{figure*}

The resulting 12 measured maps $U^\text{meas}$  are nearly unitary within error:
over all 12 circuits the maximum value of $D_{jk}{=}\lvert U^\text{meas}_{jk} U^{\text{meas}\dagger}_{jk} - \mathbbm{1}_{jk}\rvert$ was 0.050, with a mean of 0.010,  and on average $D_{jk}$ differed from zero by just 1.6 standard deviations, as determined through Monte Carlo analysis using our uncertainties in $r^\text{meas}_{jk}$ and $\phi^\text{meas}_{jk}$. Those uncertainties were derived directly from the measured variance in output power ratios and relative phase respectively taken over multiple trials.
Comparison of  the measured matrices $U^\text{meas}$ to $U^\text{circ}$ in Eq.~\eqref{eqn:Ucirc} allows nearly direct determination of the net phase $\phi_N$; notably the values of $\phi^\text{meas}_{jk}$ are consistent with $U^\text{circ}$ to within error.  The splitting parameters $\theta_n$, $n \in \{1,...,4\}$ can be determined from $r^\text{meas}$ using numerical optimisation.  The results for $F_p$, $F_m$, $\phi_N$, and $\theta_n$ for all 12 measured candidate circuits are shown in Fig.~
3.  Notably the fidelity values shown were calculated directly from the measured matrices $U^\text{meas}$, but agree to within error with the values calculated from the measured phase and reflectivity deviations in the manner depicted in Fig.~
2. The relatively higher variance in $\theta_3$ is due largely to the increased sensitivity of longer couplers to slight index mismatches, as explained in Sec.~\ref{sec:devfab}; BS3 has a coupling region almost six times the length of BS1 and BS2 in order to achieve the required internal phase shift.

For the best measured device, the mode and process fidelities determined were $F_m(U^\text{meas}, U^\text{\hcz{}})=0.931 {\pm} 0.001$ and $F_p{=}0.680 {\pm} 0.005$ respectively; the net internal phase found was $\phi_N{=}-0.346{\pm} 0.013$, and the splitting parameter deviations determined were $\Delta\theta_1{=}0.087 {\pm} 0.002$, $\Delta\theta_2{=}0.083 {\pm} 0.003$, $\Delta\theta_3=-0.065 {\pm}0.003$, and $\Delta\theta_4{=}0.337 {\pm} 0.002$. As an illustration these parameter deviations are also depicted in Fig.~
2 along with their individual effects on the fidelities. For all 12 devices BS4 was erroneously fabricated with a reflectivity near 60\% rather than the ideal value of 90.8$\%$ . However we note that if $\Delta\theta_4$ had been approximately the mean of the other splitting deviations achieved, with a value of 0.08, the process fidelity calculated according to those parameter errors would have been $F_p{=}0.882$ and the mode fidelity $F_m{=}0.984$.

\subsection{Verification using two-photon interference.}

Full operation of these circuits as gates using spontaneous parametric down-conversion (SPDC)---the current state-of-the-art in generating multiple single photons---requires a six-photon output state where two serve as triggers. This is to avoid heralding false positives due to the probabilistic nature of SPDC. Given the $\nicefrac{2}{27}$ gate success probability and our loss of at least 1.8~dB per waveguide, this would result in a success probability of less than 0.015 per six-photon input. With current six-photon SPDC generation capabilities~\cite{Prevedel2009} we would thus expect a success rate of less than 3~mHz, necessitating prohibitively long integration times and low signal-to-noise for \emph{e.g.} quantum process tomography, thereby limiting conclusions regarding actual gate fidelity.

However, despite requiring four input photons for full operation, the \hcz{} circuit relies only on fourth-order interference effects (in field), \emph{i.e.} two-photon quantum interference; any higher-order interference effects between the four input photons can only result in error terms where the control and target qubits along with the two ancillary modes do not output exactly one photon each. In order both to confirm the ability of the circuits to support high-visibility quantum interference and to verify the results of their coherent characterisation, we therefore measured the visibility of two-photon quantum interference in the best-performing circuit for all possible input-output mode combinations. These measured visibilities are compared against both predictions from our classical characterisation and the ideal \hcz{} circuit visibilities in Fig.~4.
\begin{figure*}
\begin{center}
\includegraphics*[width=0.99\textwidth]{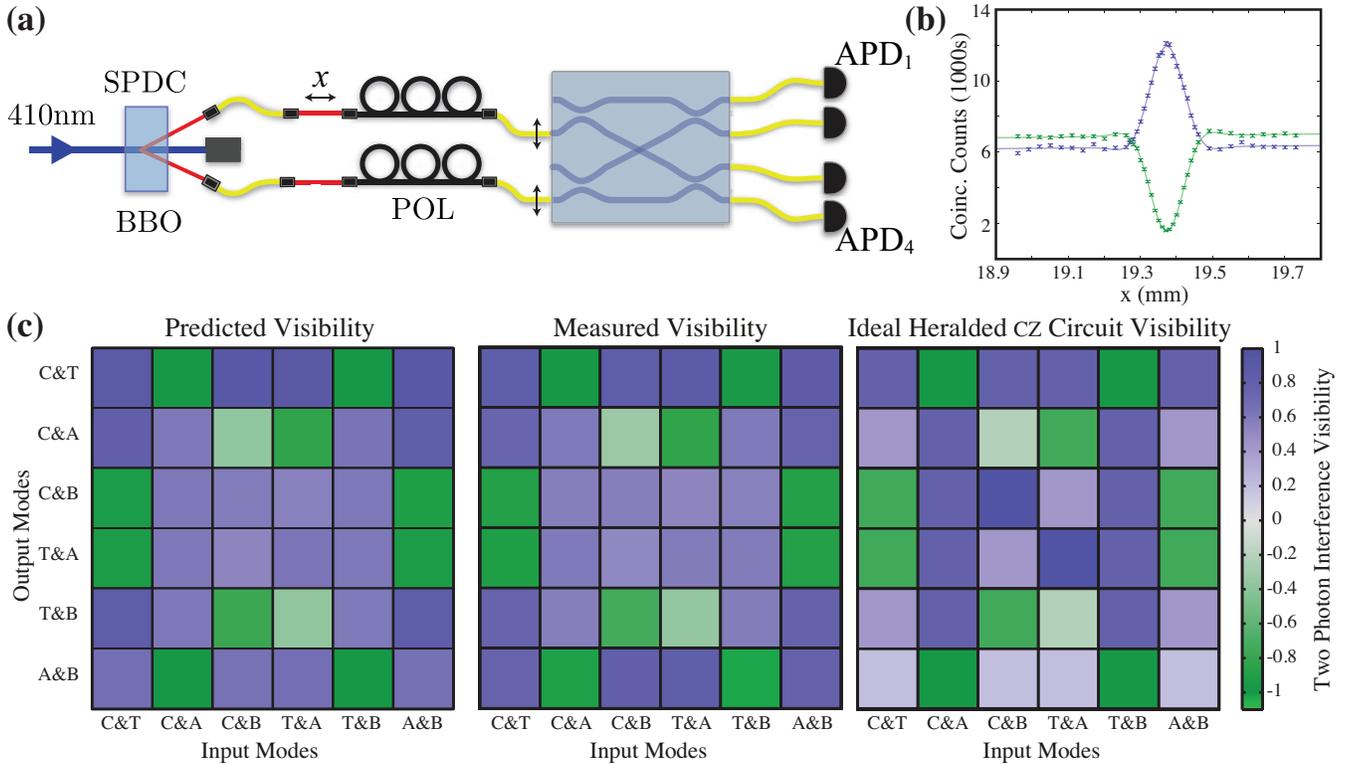}
\end{center}
\caption{
(a) Setup for measuring two-photon interference. Degenerate photon pairs at $820$~nm are created via spontaneous parametric down-conversion (SPDC) in a nonlinear $\beta$-barium-borate (BBO) crystal pumped by a $410$~nm frequency-doubled Ti:sapphire laser. Manual polarisation controllers (POL) enable alignment of SPDC polarisation with the axes of polarisation-maintaining fibers coupled to the test device. We detect photons in coincidence using avalanche photo diodes (APD). (b) Representative observed non-classical interference patterns, showing anti-coalescent and coalescent interference for two different output mode combinations as well as fits to the data with Gaussian and sinc components. The error bars shown stem from Poissonian counting statistics. (c) Two photon interference results for our best \hcz{} circuit. We compare predictions (left) from the coherently-characterised circuit against measured two-photon interference visibilities (centre). The right panel shows visibilities for the ideal circuit $U^\text{\hcz{}}$; most of the difference between this panel and the other two is due to the deviation in our best circuit from the ideal reflectivity for BS4 and unwanted net phase $\phi_N$, as shown in Fig.~
3. For the residuals after subtracting the measured visibilities from the predictions the mean and standard deviation are -0.002 and 0.061, while after subtracting the measured values from the ideal values they are -0.022 and 0.282 respectively.
 }
\label{fig:quantum_char}
\end{figure*}

The apparatus for measuring the quantum interference effects is depicted schematically in Fig.~
4(a).  We measured two-photon quantum interference visibility for all ${4 \choose 2}^{2}{=}36$ combinations of two input and two output ports.  The interference visibility $V$ is calculated as $V{=}\left(C_{\mathrm{max}}{-}C_{\mathrm{min}}\right)/C_{\mathrm{max}}$, where $C$ is the rate of coincident photon detection events as a function of the temporal delay between the input photons, and $C_{\mathrm{max}}$ and $C_{\mathrm{min}}$ are calculated from a fit to the data as shown in Fig.~
4(b).

The measured visibilities are shown in Fig.~
4(c), along with those predicted from $U^{\text{meas}}$ as determined via coherent characterisation, and the visibilities for an ideal \hcz{} circuit. 
The mean of the residuals after subtracting the measured visibilities from the predictions is only -0.002 with a standard deviation of 0.061.
Perhaps a better comparison is achieved by numerically calculating the unitary $U^\text{vis}$ which would yield the minimum root-mean-square difference from the measured visibilities; this unitary has mode fidelities of $F_m(U^\text{meas},U^\text{vis})=0.983$ and $F_m(U^{\text{\hcz{}}},U^\text{vis})=0.931$ with the measured circuit and the ideal \hcz{} respectively. The small differences between predicted and measured visibilities can be attributed largely to three factors: polarisation non-degeneracy between the interfering photons in the FLDW circuit; the slightly differing spectra of the SPDC photons and the laser diode used for the coherent characterisation; and the effects of higher-order SPDC terms. 

\section{Conclusions}

Along with further improvements in photon sources and detection, heralding will be required to concatenate multiple entangling LOQC gates and thus enable more complex quantum computations and simulations. We have demonstrated that integrated waveguide arrays, particularly using femtosecond laser-writing, are capable of generating the required multimode interference circuits with both high fidelity and excellent quantum interference, and allow simple implementation of mode crossover elements and $\pi$ phase shifts. However further careful engineering will be required to precisely achieve the desired beamsplitter ratios and to avoid undesired phase accumulation. This study has outlined the challenges both experimentally and theoretically in achieving waveguide circuits with high operational fidelities.

The quantum process fidelity of candidate circuits can be calculated from known fabrication tolerances or classical characterisation results using the Jamiolkowski isomorphism, and this metric has proven to be more sensitive and useful than mode fidelity for assessing such circuits. However any circuit imbalance will lead to error terms outside the computational subspace wherein two photons exit in either the control or target mode. The precise effects of such coherent error terms when multiple gates are concatenated, as well as possibilities for their correction or mitigation, could be a fruitful avenue for future investigation.

\section*{Acknowledgements}
This research was supported in part by the Australian Research Council Centre of Excellence for Ultrahigh bandwidth Devices for Optical Systems (CE110001018), the Centre of Excellence for Quantum Computation and Communication Technology (CE110001027), and the Centre of Excellence for Engineered Quantum Systems (CE110001013). AF is supported by an Australian Research Council Discovery Early Career Research Award (DE130100240), and AGW by by a UQ Vice-Chancellor's Senior Research Fellowship.


\end{document}